# Enabling Cross-Layer Reliability and Functional Safety Assessment Through ML-Based Compact Models

Dan Alexandrescu
IROC Technologies
Grenoble, France
dan@iroctech.com

Aneesh Balakrishnan
IROC Technologies,
Grenoble, France
Tallinn University of
Technology, Tallinn 12618,
Estonia
aneesh.balakrishnan@iroctech.com

Thomas Lange
IROC Technologies
Grenoble, France
Dipartimento di Informatica e
Automatica, Politecnico di
Torino, Torino, Italy
thomas.lange@iroctech.com

Maximilien Glorieux
IROC Technologies
Grenoble, France
maximilien.glorieux@iroctech.com

*Abstract*—**Typical design flows are hierarchical and rely on assembling many individual technology elements from standard cells to complete boards. Providers use compact models to provide simplified views of their products to their users. Designers group simpler elements in more complex structures and have to manage the corresponding propagation of reliability and functional safety information through the hierarchy of the system, accompanied by the obvious problems of IP confidentiality, possibility of reverse engineering and so on. This paper proposes a machine-learning-based approach to integrate the many individual models of a subsystem's elements in a single compact model that can be re-used and assembled further up in the hierarchy. The compact models provide consistency, accuracy and confidentiality, allowing technology, IP, component, sub-system or system providers to accompany their offering with high-quality reliability and functional safety compact models that can be safely and accurately consumed by their users.**

I. INTRODUCTION

High quality, reliable and safe electronics requires massive cooperation and exchanges across all the partners of the supply, design and manufacturing flow. A huge quantity of information, addressing functional and extra-functional qualities must be produced accurately, exchanged without loss of fidelity and consumed as intended.

One of the most self-evident examples of such flow of information is the typical foundry>designer>integrator process. The technology provider prepares a complex Process Design Kit that includes technological data, simulation models, design rules information, primitives and possible standard cell libraries from the foundry or a library vendor. Designers make use of this information during the preparation of cells, IP blocks and ultimately components integrating the technology elements. Components are used by system integrators on boards, sub-systems and systems. Some information (such as recommended supply voltage) will be valuable through the whole flow while other data can be consumed in one design stage to ensure the fulfilment of specific requirements.

Design paradigms, workflows, practices and expectations are progressing continuously. While in the past a typical ASIC design process was concerned by the paramount trinity of area/frequency/power, todays requirements are presented as a vast set of functional and extra-functional specifications. Functional Safety (FuSa) and Reliability (Rel) requirements are increasingly present because of implicit or explicit customer expectations or, more stringent, because of current or upcoming formal standards such as IEC 61508 [1], ISO 26262 [2] and others. These standards demand the calculation and presentation of functional safety and reliability metrics at system-level, in quantitative and qualitative terms. However, computing the Failure Rate of a system to make sure that it fulfils the <10 FITs requirement for an ASIL D product involves data that has been produced by at least three individual entities and transformed by tens of engineers (or even companies) during the design flow. There is a huge potential for loss of fidelity, data misuse, omissions and translation errors. In addition, much of this information is highly proprietary and the transmission of the information from one partner of another can be parasitized by restrictions and disclosure limitations. Some of the data can also facilitate possible reverse engineering or at least disclose critical design information that was intended to be kept secret.

In this paper we present a uniform methodology to evaluate reliability and functional safety metrics based on using compact models build with Machine-Learning (ML) methods. The compact models can be used at any design abstraction level. The compact models of a design element (standard cell, IP, block, component, sub-system or system) at a given hierarchical level can be combined through ML techniques in a single compact model that can be used for the next design stage or user. This approach ensures that relevant parameters are retained and impactful during the design flow and that the contribution of the various design elements is well represented at any calculation stage. Additionally, it may provide a way to obscure technology and design information, safeguarding critical IP, while still

This work was done as part of the RESCUE project that has received funding from the European Union's Horizon 2020 research and innovation programme under the Marie Skłodowska-Curie grant agreement No. 722325 equipping the users at any design stage with the information they need for their work.

The paper is organized as follows: Section II introduces the motivation for this work, Section III presents the proposed approach, Section IV shows a worked example and it's followed by the Conclusion section.

## II. MOTIVATION

As an example of a metric calculation flow and the associated perils and works, let's consider the evaluation of Single Events Effects (or Soft Errors or Transient Faults) to the failure rate of a system:

Firstly, the technology provider (foundry), in a possible collaboration with a library vendor must provide technological (raw) event rate for the various technology elements (standard cells, memory blocks, analog IP such as PLLs and so on) that its customer uses. Sequential cells can be affected by Single Event Upsets (SEU) with rates that depend on the cell state, voltage, temperature and so on. Single Event Transients in combinatorial cells can also depend on the cell fanout. Single or Multiple Bit/Cell Upsets (SBU/MCU/MBU) can impact the data stored in memory blocks and their occurrence rate can depend on physical implementation (aspect, column muxing, scrambling) or design choices such as error management schemes. Ideally, all this data deserves to be captured in a detailed, high fidelity format that accurately reproduces the error rate of the researched event according to a reasonable set of parameters. There is massive precedence for this approach: today's PDK contain large databases, multi-index tables and process information in a variety of formats. While this may happen one day, the current State-Of-The-Art in terms of SEE technological information delivery consist in PDF test reports or a spreadsheet with limited summary information.

The component or IP designer must use this information as an input of its own event rate calculation methodology. As an example, the unit SEU rates indicated by the technology provider must be annotated to each design flip-flop and de-rated according to the role of the flip-flop in the design:

$$SEQ\ Fail\ Rate = \sum_{Flip-Flop} Raw\ SEU\ Rate \cdot \prod_{DR} De-Rating \quad (1)$$

Memory SBU/MCU data is tailored according to each instance specifics and the impact of possible error management schemes is integrated.

$$MEM\ Fail\ Rate = \sum_{Memories} \begin{cases} Raw\ SBU\ Rate \cdot Size\ if\ no\ ECC \\ Raw\ MBU\ rate \cdot Size\ if\ ECC \end{cases} \quad (2)$$

The end goal is to calculate at the IP or design level the required overall reliability or functional safety metrics. While in some rare circumstances, the final deliverable is a number or a limited set of numbers, any actual high-level metric will be dependent on a variety of parameters. Firstly, some parameters (such as voltage, temperature) inherited from the underlying technological layer will certainly impact the transient fault error rate. It may still be feasible at this stage to provide indexed tables containing the necessary data for the possible combinations of a few given parameters. However, in more complex cases, this is not sufficient, as the design can be configured differently or used in various scenarios, affecting the exhibited failure rate. Complex ASICs can be configured to fulfill different function modes. Design settings (speed, number of active cores, buffer or cache sizes, memory modes) can be set according to each application requirements. Internal features (error detection and/or correction, safety mechanisms) can be activated or not with a direct impact on the design failure rate. CPU-based designs can show various failures with different manifestations and rate that depend on how the CPU core(s) are used by a given application.

The same design can have multiple physical implementation that can also impact reliability. As an example, the usage of packaging materials with different alpha emissivity rates will strongly impact (from x1 to x1000) the final SEU rate.

In conclusion, the set of parameters affecting the FuSa/Rel metrics of a design can be large and impactful. The manufacturer will usually have difficulties in packaging and transmitting this information to the user. There are several options:

- A maximum, worst-case metric. This way, the actual failure rate is guaranteed to be lower than the indicated value. This approach can penalize some applications or systems and can lead to overengineering in trying to manage the failure rate.

- A recommended utilization scenario that leads to a nominal, favorable error rate. This scenario is usually implemented through a "Safety Manual" where the manufacturer describes his vision on how the component should be used in order to fulfil safety goals and to lead to the desired failure rate

- An analytical method where the provider can explicitly describe a function to calculate the actual failure rate according to the implementation. As an example, assuming that the system includes memory instances, then their contribution to the overall failure rate can be described similarly to the equation (2). However, this approach presents the drawbacks of disclosing to the user in a clear format, internal circuit information or raw technology data which can be critical pieces of IP or know-how.

Regardless of how the information is provided, the user of the design will use this component in its own system and will need to evaluate the failure rate for its own implementation. Moreover, a system integrator will typically use many individual components from various providers. Translating, adapting and integrating various reliability data is very challenging. Finally, the system integrator will have to deliver a final product with ideally, a clear set of metrics that can be compared to the requirements. However, even the system reliability metrics can depend on parameters that can vary according to the usage. The same exact product can show different failure rates when used in a datacenter, a car, an airplane or a satellite. The provider needs to express the metrics of its products according to a set of parameters and has to provide the user with a metric model with a set of parameters that are relevant for the given abstraction level.

The various collaborations between the providers and users of technology, components and systems need to be supported by adequate tools, preferably developed for reliability and functional safety efforts. While more specific tools exist, standard spreadsheet tools are widely deployed and used. Data import and export from these tools is inadequate and prone to errors.

Recent standardization efforts from IEEE [3] and Accellera [4] aims at defining an exchangeable interoperability format for functional safety analysis and functional safety verification activities. A format for Reliability exchanges has been proposed previously [5], providing a solution to suppliers and consumers to exchange reliability information in a consistent fashion and to use this information to construct accurate reliability models. All these efforts stress the importance of building models hierarchically and according to the design abstraction level. Accordingly, an uniform, universal method to model functional safety or reliability metrics at any design stage or abstraction level and more importantly, combining the information available at the current level to benefit the next efforts in the pipeline would be opportune and useful.

Using Machine Learning techniques to build such a model has several advantages. Machine Learning algorithms are known to efficiently learn even complex relationships. Models can be built from various types of input data, such as tables or functions. Therefore, it is suitable at any abstraction and hierarchical level. Several aspects can be combined in a single model, which makes it compact and easy to use by the user in the next design stage. Additionally, the representation of the Machine Learning model can be fundamentally different to the original representation and thus, obscure critical technology and design information.

### III. PROPOSED METHODOLOGY

The proposed approach is conceptually straightforward and easy to implement. The methodology relies on compact models build using Machine-Learning techniques and trained using reference data provided by State-of-the-Art (SOTA) methods. The overall FuSa/Rel metrics for the current level can be then calculated and used as reference data to train a compact ML model. The elaborated ML model can then fuel the analysis efforts required for the next design stage or upper hierarchical level or abstraction and can be shared with the corresponding engineer or user.

The phases of this methodology are described below:

*Phase I – Data collection*

Let's assume that the current hierarchical level or design abstraction is a collection of elements (or components). Each element has one or multiple attached FuSa/Rel metrics and models. The models can be analytical, data or ML. Each individual metric can have a collection of input parameters:

$$Metric_i = f(par_{i,0}, par_{i,1}, par_{i,2}, \quad \ldots par_{i,n(i)}) \quad (3)$$

*Phase II - Integration*

At the current level, SOTA approaches will allow us to combine the existing data of the various elements in overall, top-level FuSa/Rel metrics. This metric will depend on all of the parameters of the element metrics, including options, parameters and choices that are applicable at this design level. Obviously, the parameter list can be simplified by optimizing the parameters that are the same or equivalent. As an example, a supply voltage parameter can be applicable to several elements.

$$Metric_{overall} = \sum_i Metric_i = f(par_0, par_1, par_2, \ldots par_m) \quad (4)$$

*Phase III – Compact ML Model Elaboration*

The equations that composes the overall metric can be then exercised over the validity range of the parameters. The overall metric can also be a collection of data valid over a specific range. The collected data will be then used to train a ML model that will accept as inputs the aggregated set of parameters and will extend to the area covered by the available dataset. Once trained, the ML model is expected to have good if not perfect accuracy over the valid range and is an ideal surrogate or replacement for the discrete overall metric.

The main goal of the training is to create a model which accurately represents the given metrics function or collection of data. The model should be able to accurately recall the trained values and depending on the learned metric, the model should also be able interpolate and extrapolate the data.

In comparison to classical Machine Learning applications this approach is a bit different. Usually the input parameter range is limited and known and for most cases more training data can be generated or gathered. In this way the data to train the model can be increased to improve the accuracy until a specified target is met.

*Phase IV – Packaging and Reuse*

The elaborated ML model can be then provided to the next engineer or user that can start applying this methodology on the next hierarchical level, flow stage or design abstraction level.

The presented methodology shows distinct advantages. The approach is uniform regardless the location in the design flow or design hierarchy. The ML models are compact, the training is not computationally intensive and implementations are available in a variety of language and programming frameworks. In many cases the trained ML model will hide or obscure sensitive design or technology information.

### IV. WORKED EXAMPLE

In this example, we are addressing the calculation of a system Transient Fault / Soft Error Rate. For the purpose of the demonstration, we will use simple, naive equations for modeling the error rates and aggregating the contribution of the various elements. Firstly, let's introduce the following functions for the calculation of the various Soft Error Rates (indicated in FITs/MegaBit or MegaCell):

$$RawSER_{Seq}(V_{dd}) = 100 \cdot (1 + (1.2 - V_{dd})) \quad (5)$$

(A Flip-Flop has a 100 FITs/Mb at the nominal 1.2V supply voltage; SER decreases at higher voltages and increases at lower). Vdd means supply voltage

$$RawSER_{Comb}(V_{dd}, PW) = 50 \cdot (1 + (1.2 - V_{dd})) \cdot \frac{50}{PW} \quad (6)$$

(A combinatorial cell can exhibit a spectrum of Single Event Transients with decreasing event rate for larger, longer events. We will limit the minimal PW at 10ps which is the shortest transient that the selected technological process can propagate). PW means Pulse Width (in ps).

The presented models for the error rate of Sequential and Combinatorial are analytical models (simple equations), but they can also be delivered as ML models accepting a voltage and a pulse width parameter.

The Single Event Effect rate for natural working environments is dominated by the contribution of SEEs caused by **alpha** particles emitted by impurities in the packaging materials and **neutrons** caused by the interaction of high-energy particles with the atmosphere. Both contributions depend on a large number of parameters and it can be difficult to provide a model that integrates the effect of all the parameters.

As an example, the **neutron**-induced SEE rate depends on many factors, including physical location, altitude, solar activity, shielding and so on. Following the approach from [7], let's focus on altitude and cutoff (dependent on location), ignoring solar modulation. In this case, the actual neutron flux (NF) at a given location can be expressed as follows:

$$NF = NF_{ref} \cdot GRF \cdot e^{-\frac{A-A_{ref}}{L}} \qquad (7)$$

, where NF$_{ref}$ is the the neutron flux at the reference location (New York = 14 n/sq. cm/h). L is the flux attenuation length for neutrons in the atmosphere (~148 g/cm2). Finally, A is the areal density of the location of interest, Aref is the areal density of the reference location.

$$A = 1033 \cdot e^{-0.03813 \cdot \left(\frac{a}{1000}\right) - 0.00014 \cdot \left(\frac{a}{1000}\right)^2 + 6.4 \cdot 10^{-7} \cdot \left(\frac{a}{1000}\right)^3} \qquad (8)$$

While these factors can be described analytically and integrated in the various models, the GRF represents the Geomagnetic Rigidity Factor and varies according to the geographical position. As an example, values of geomagnetic vertical cutoff rigidity used to calculate the relative neutron flux were provided by the Aerospace Medical Research Division of the Federal Aviation Administration's Civil Aerospace Medical Institute. The cutoff data were generated by M.A. Shea and D.F. Smart using the International Geomagnetic Reference Field for 1995 [8][9]. Therefore, the actual GRF values can only be provided as a table of data indexed according the longitude and latitude. This cause a number of issues, including the need to interpolate between the available sparse, low granularity data and the difficulty to integrate tabular data in a compact model.

Machine-Learning Models can cope very efficiently with these difficulties. Firstly, the training can be done on any type of data, analytical, tabular with any type of data representation for the input parameters: linear (for the altitude) or specific (geographical coordinates). Moreover, it will also be able, provided that an adequate model is used, to interpolate GRF data between the locations provided in the tables, allowing the approximate calculation of the GRF for any location.

The neutron-induced error rate is provided as base value for the reference setting (New-York, sea-level) that needs to be multiplied by an acceleration factor calculated according to the actual location and altitude. As an example, the following tables show the acceleration factors for a selection of altitudes and location. The final acceleration factor can be calculated by multiplying the appropriate location factor with the desired altitude factor.

TABLE I. ALTITUDE-DEPENDED NEUTRON DATA

| *Altitude* (feet) | *Altitude* (m) | Neutrons flux (n/sq.cm hour) | Neutrons flux (n/sq.cm second) | Neutrons flux (relative to sea level) |
|---|---|---|---|---|
| 0 | 0 | 14.0 | 0.003889 | 1.0 |
| 1000 | 304.8 | 18.2 | 0.005056 | 1.3 |
| 2000 | 609.6 | 23.4 | 0.0065 | 1.7 |
| 5000 | 1524 | 47.6 | 0.013222 | 3.4 |
| 10000 | 3048 | 134.6 | 0.037389 | 9.6 |
| 20000 | 6096 | 668.5 | 0.185694 | 47.8 |
| 30000 | 9144 | 2001.1 | 0.555861 | 142.9 |
| 35000 | 10668 | 2993.2 | 0.831444 | 213.8 |
| 40000 | 12192 | 4147.0 | 1.151944 | 296.2 |

TABLE II. LOCATION-DEPENDED NEUTRON DATA

| *Location* | Neutrons flux (relative to sea level) |
|---|---|
| Milpitas | 0.92 |
| Colorado Springs | 4.42 |
| Bangalore | 1.02 |
| Beijing | 0.72 |
| Grenoble, France | 1.24 |

This part fulfils the Phase I of the proposed methodology.

In the following (Phase II), a simple de-rating approach is assumed to calculate the overall Error Rate of an ASIC with 1 Mbit of Flip-Flops and 10 Mbits of Combinatorial cells. The overall Error Rate can be computed as the de-rated contribution of each individual element:

$$SER_{ASIC} = \sum_{element} RawSER_{s|c} \cdot \prod_{LDR,TDR,FDR} DR \qquad (9)$$

The de-ratings applicable here are: LDR – Logic De-Rating (the probability of the fault to propagate from a logic/masking perspective), the TDR – Temporal De-Rating (the probability of the fault to arrive during a sensitive opportunity window) and FDR – Functional De-Rating (the probability for an error to become an observable failure). The Fault/Error/Failure dichotomy and the various De-Rating factors are used as presented in [6].

Simple equations or values for the applicable De-Rating factors are presented in the equations below

$$TDR_{Seq}(Freq) = \frac{Slack}{ClockPeriod} = 1 - \frac{Freq[Hz]}{2e6} \qquad (10)$$

$$TDR_{Comb}(PW, Freq) = \frac{PW}{ClockPeriod} = PW * Freq \quad (11)$$

$$LDR = 0.25 \quad (12)$$

$$FDR = 0.25 \quad (13)$$

Finally, the overall SER of the ASIC can be described as follows:

$$\begin{aligned}SER(PW, Freq, V_{dd}) \\ = \big(1Mbit \cdot RawSER_{Seq} \cdot TDR_{Seq} \\ + 10Mbit \cdot RawSER_{Comb} \cdot TDR_{comb}\big) \\ \cdot LDR \cdot FDR \quad (14)\end{aligned}$$

and can be unrolled in a clear format:

$$\begin{aligned}SER(PW, Freq, V_{dd}) \\ = \bigg(1Mb \cdot 100\frac{FIT}{Mb} \cdot \big(1 + (1.2 - V_{dd})\big) \cdot 1 \\ - \frac{Freq[Hz]}{2e6} + 10Mb \cdot 50\frac{FIT}{Mb} \cdot (1 \\ + (1.2 - V_{dd})) \cdot \frac{50}{PW} \cdot PW * Freq\bigg) \cdot 0.25 \\ \cdot 0.25 \quad (15)\end{aligned}$$

A final customization can be made to clarify the value or the value range for the Pulse Width parameter which is a technology attribute and may not make sense to or be fillable by the final user. Therefore, the ASIC provider, in agreement with the technology provider, may specify values for the PW according to the working environment. Let's consider a neutron environment (ground applications) with a typical PW of 50ps, Alpha particles – 10ps and Heavy Ions – 100ps.

It is obvious that the overall SER equation can disclose unit technology data (FIT rates per Mb, typical pulse widths), design structure (1Mb of FF and 10Mb of combinatorial cells) or knowledge (such as calculation of de-rating factors).

The next step (*Phase III – Compact ML Model Elaboration*) consists in exercising the overall SER equation over the range of valid values for the environment, frequency and voltage. The results of this exploration are presented in the following table:

TABLE III. SER VALUES TABLE

| Sample | Parameters ||||  SER [FIT] |
|---|---|---|---|---|---|
| | Voltage [V] | Environment | Frequency [MHz] | Location, Altitude | |
| 0 | 0.8 | n | 0.1 | NYC, sea-level | 238.551 |
| 1 | 0.8 | n | 0.2 | NYC, sea-level | 467.927 |
| 2 | 0.8 | n | 0.3 | NYC, sea-level | 697.303 |
| 3 | 0.8 | n | 0.4 | NYC, sea-level | 926.679 |
| ... | ... | ... | ... | ... | ... |
| 896 | 1.5 | HI | 1.7 | N/A | 1954.28 |
| 897 | 1.5 | HI | 1.8 | N/A | 2068.97 |
| 898 | 1.5 | HI | 1.9 | N/A | 2183.66 |
| 899 | 1.5 | HI | 2 | N/A | 2298.35 |

*Phase III* continues with the training of the ML model. In this work we are interested by the applicability of different ML models for the intended usage. Accordingly, several ML regression methods have been evaluated: Linear Models (Linear and Ridge), Kernel Ridge Regressor (with Linear, Polynomial, RBF, Sigmoid Kernels), Decision Tree Models, Neighbors-based Models (K-Nearest and Radius), Support Vector Machine Models (Linear SVR, SVR with various kernels, NuSVR), Multilayer Perceptron Neural Networks.

All models have been trained with 60% of the data set, the train data set. After the training, the models are exercised over the full permitted parameters range. This allows, on the one hand, to measure the accuracy of the model to recall values which were already in the training data set. On the other hand, by testing the model with data not used for training it is evaluated if the model is able to interpolate and extrapolate the given data. The performance of the models is measured by comparing the predicted values against the reference values and calculate the following metrics: "MAE - Mean Absolute Error", "MAX - Max Absolute Error", "RMSE – Root Mean Squared Error", "EV - Explained Variance", "$R^2$ - Coefficient of Determination".

The following table presents the results for the most accurate and promising models:

1. Ridge Regression with Polynomial Kernel
2. Ridge Regression with RBF Kernel
3. k-Nearest Neighbors Regression
4. Support Vector Regression with Polynomial Kernel

TABLE IV. TOP ML MODELS PERFORMANCES

| ML Model | Data Set | ML Model Error/Correlation Measurements |||||
|---|---|---|---|---|---|---|
| | | MAE | MAX | RMSE | EV | $R^2$ |
| 1 | Train | 1.688e-06 | 1.199e-05 | 2.119e-06 | 1 | 1 |
| | Test | 1.695e-06 | 1.331e-05 | 2.106e-06 | 1 | 1 |
| 2 | Train | 2.716e-05 | 2.545e-04 | 3.530e-05 | 1 | 1 |
| | Test | 2.907e-05 | 6.609e-04 | 4.421e-05 | 1 | 1 |
| 3 | Train | 0 | 0 | 0 | 1 | 1 |
| | Test | 21.91 | 195.2 | 30.98 | 0.99 | 0.99 |
| 4 | Train | 1.787e-02 | 6.830e-02 | 2.233e-02 | 1 | 1 |
| | Test | 1.756e-02 | 6.243e-02 | 2.195e-02 | 1 | 1 |

The results show that the presented Machine Learning models are able to accurately recall the values from the train data set. The error metrics MAE, MAX, and RMSE are very low and the correlation like metrics EV and $R^2$ are 1 for most of the models. The Predicting the test data set show similar good results, which means that the models are also able to interpolate and extrapolate the given data. The k-Nearest Neighbors regression is able to perfectly recall the data the data used for training the model, due to the nature of its algorithm. However, predicting new values from the test data set, which were not used for training, shows the models weakness. The inter- and extrapolation capability is less good in comparison to other models.

Fig. 1 shows the graphs representing the model prediction error for the train and test data set, as well as the correlation between the actual and predicted values when the Ridge

Regression with polynomial Kernel is used. In comparison Fig. 2 the results for the k-Nearest Neighbors regression is shown. The graphs show as well very clearly, that the k-Nearest Neighbors regression is perfectly able to recall the training data but less efficient when predicting the test data.

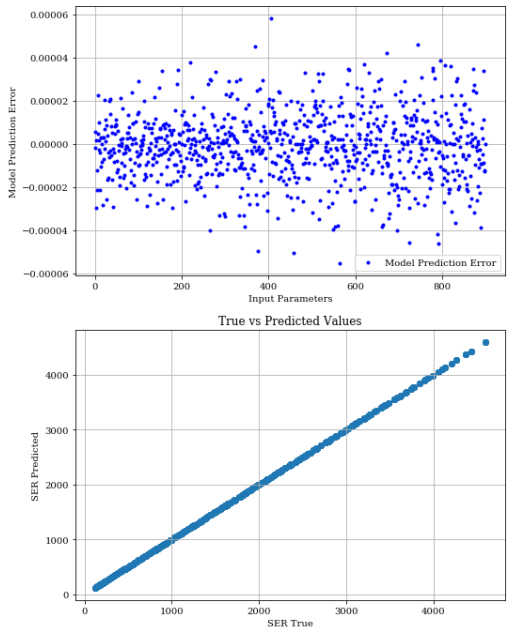

Fig. 1. Results of predicting the train and test data set by using Ridge regression with polynomial kernel.

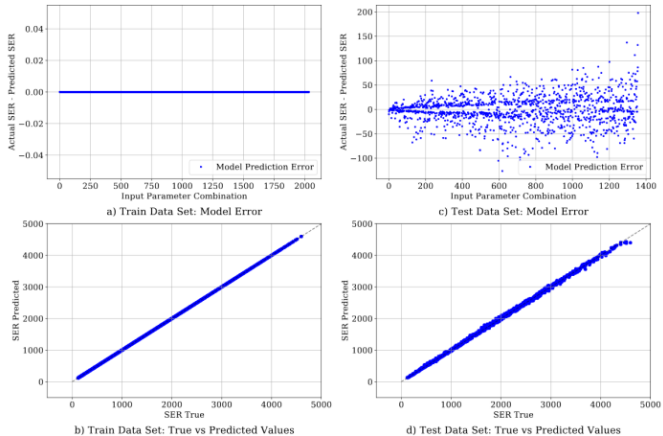

Fig. 2. Results of predicting the train and test data set by using k-Nearset Neighbors regression.

The results show that certain ML models are able to learn the given metric function and it is a good approach for the matching of extensive reference data sets. Depending on the data or function to learn other models can be less effective. As seen in the example they might only be appropriate to recall the data but not to inter- and extrapolate it. Other models might be less effective in general. This means for the given data several models need to be considered and evaluated.

TABLE V. LINEAR REGRESSION MODEL

| ML | ML Model Error/Correlation Measurements | | | | | Time (s) |
|---|---|---|---|---|---|---|
| | MAE | MAX | RMSE | EV | $R^2$ | |
| 1 | 152.917 | 544.768 | 204.089 | 0.96 | 0.96 | 0.0025 |

The graphical representation of this model is also telling (Fig. 3)

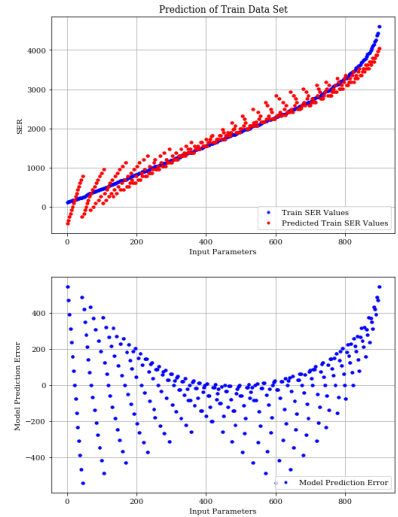

Fig. 3. Example of results of the Linear Regressor Model

Because of the limited and simple training data set, the training of the ML models is fast and doesn't require computationally intensive resources. A single-shot execution of a trained models is very fast for most models, with execution time inferior to the nanosecond.

## V. CONCLUSIONS AND PERSPECTIVES

In this paper, a methodology has been proposed to help experts to approach the complexity of hierarchical modeling of reliability and functional safety metrics. Furthermore, the presented approach allows the use, elaboration and distribution of compact ML models in an uniform and systematic manner, minimizing both human and CPU efforts while maintaining high accuracy and fidelity.

While the approach has been validated and seems to work effectively and efficiently on a modest example, further works will have to consider more complex systems with expanded sets of parameters and with more, non-linear fault models.

Finally, this approach can be used to integrate the effect of multiple failure mechanisms (such as Total Ionizing Dose – TID, circuit aging) that can contribute to an overall applicative failure rate. The effects of such mechanisms depend on generic (such as circuit structure, test-vector/workloads, PVT …) and specific (total and rate of received dose, age of the circuit …), The authors intend to to train ML models for a combination of Transient Faults/Soft Errors, TID and Aging Effects using the same fundamental platform of parameters plus a limited set of effect-specific parameters. The trained ML models will be able to quickly evaluate a large variety of effects, encapsulating very useful reliability and functional safety data in a compact and efficient solution that can be used and reused further down the design and manufacturing flow.